# BAMCloud: A Cloud Based Mobile Biometric Authentication Framework

Farhana J. Zareen*, Kashish A. Shakil*, *Student Member, IEEE,* Mansaf Alam, *Member, IEEE,* Suraiya Jabin and Shabih Shakeel

*Abstract*—With an exponential increase in number of users switching to mobile banking, various countries are adopting biometric solutions as security measures. The main reason for biometric technologies becoming more common in the everyday lives of consumers is because of the facility to easily capture biometric data in real time, using their mobile phones. Biometric technologies are providing the potential security framework to make banking more convenient and secure than it has ever been. At the same time, the exponential growth of enrollment in the biometric system produces massive amount of high dimensionality data that leads to degradation in the performance of the mobile banking systems. Therefore, in order to overcome the performance issues arising due to this data deluge, this paper aims to propose a distributed mobile biometric system based on a high performance cluster Cloud. High availability, better time efficiency and scalability are some of the added advantages of using the proposed system. In this paper a Cloud based mobile biometric authentication framework (BAMCloud) is proposed that uses dynamic signatures and performs authentication. It includes the steps involving data capture using any handheld mobile device, then storage, preprocessing and training the system in a distributed manner over Cloud. For this purpose we have implemented it using MapReduce on Hadoop platform and for training Levenberg-Marquardt backpropagation neural network has been used. Moreover, the methodology adopted is very novel as it achieves a speedup of 8.5x and a performance of 96.23%. Furthermore, the cost benefit analysis of the implemented system shows that the cost of implementation and execution of the system is lesser than the existing ones. The experiments demonstrate that the better performance is achieved by proposed framework as compared to the other methods used in the recent literature.

*Index Terms*— authentication, BAMCloud, biometrics, Cloud computing, machine learning

## I. INTRODUCTION

THE recent upheaval of touch screen technology is providing a viable environment for implementation of mobile biometric system. Biometric system is used to authenticate an individual. Dynamic signature is one such biometric modality which is used in establishing identities as well as authentication and is socially and legally accepted [1]. Dynamic signatures use structural and behavioural characteristics that are exhibited by an individual while signing his/her name. They are recorded using a digitizer that captures a number of dynamic properties of a signature such as pen-pressure, time stamp, co-ordinates, velocity, acceleration etc.

Dynamic signature authentication is a relatively new approach for user-validation which can replace passwords or tokens, that may be forgotten or stolen [2]. One problem in the implementation of signature biometric system in mobile environment is that, a reasonable number of signature samples have to be captured from each user so that the system can be trained properly, along with that a number of forged samples should also be captured to train the system. Each signature sample contains a lot of information e.g. pressure, tilt angle, velocity etc. and there can be millions of users whose signatures are to be stored as templates. For example in a developing country like India with its population being the second highest in the world. If enrollments are to be done for a unique identification card (Aadhaar, the most ambitious project of Indian government with objective to collect the biometric and demographic data of Indian residents, store them in a centralized database, and issue a 12-digit unique identity number (uid) called Aadhaar to each resident), even if we cover half the country we will end up doing 4000000 (number of enrollments per day) x 12 (number of entries) x 500000000 (number of existing uids) x 12 (number of entries per uid) matches of biometric samples in a day [3]. It is nearly impossible to be done by a single processor of a hand-held device and therefore a data management issue arises which

This paragraph of the first footnote will contain the date on which you submitted your paper for review. It will also contain support information, including sponsor and financial support acknowledgment. For example, "This work was supported in part by the U.S. Department of Commerce under Grant BS123456".

The next few paragraphs should contain the authors' current affiliations, including current address and e-mail. For example, F. A. Author is with the National Institute of Standards and Technology, Boulder, CO 80305 USA (e-mail: author@ boulder.nist.gov).

Farhana Javed Zareen is with the Department of Computer Science, Jamia Millia Islamia, New Delhi 110025. E-mail: farhanazareen@yahoo.com.

Kashish Ara Shakil is with the Department of Computer Science, Jamia Millia Islamia, New Delhi 110025. E-mail: shakilkashish@yahoo.co.in.

Mansaf Alam is with Department of Computer Science, Jamia Millia Islamia, New Delhi 110025. E-mail: malam2@jmi.ac.in

Suraiya Jabin is with Department of Computer Science, Jamia Millia Islamia, New Delhi 110025. E-mail: sjabin@jmi.ac.in

Shabih Shakeel is with Institute of Biotechnology, University of Helsinki, Finland. E-mail: shabih.shakeel@helsinki.fi

*Equal Contributions

needs to be resolved. Moreover even though if a system is able to handle such kind of data, there is always a performance degradation arising due to computationally intensive tasks.

Though work has been done in other fields of biometric authentication for large samples such as face recognition [4], [5] using all-pairs which is a Cloud based data abstraction for data intensive computing tasks [6], animal identification [7], brain mapping [8] and finger print recognition system [9] but work in the field of signature biometrics with large datasets is still at its nascent stage. Since signatures are the most socially accepted authentication mechanism therefore with growing size of database this field must be given more attention. In recent times, smart phones and tablets have become very popular in accessing all kind of services and information. Thus, implementing biometric system in mobile devices has become an attractive target [10]. Use of mobile phones and handheld devices acts as a cheap solution in comparison to other available biometrics. Though online signature verification is quite a popular technique but use of mobile devices in this area is still not tapped. Hence, in the proposed approach mobile and handheld devices have been used to capture samples since it provides a cost efficient solution.

In order to solve data management problem and thereby performance issues arising due to large number of signature samples, distributing the processing and storing the data over a Cloud comes up as an inherent solution. The target of BAMCloud is to provide a scalable [11] and cost efficient technique for handling and processing growing data on Cloud without compromising on the accuracy of the biometric system. Also it is to be noted that to match the requests which include authentication and enrollment in a biometric system, the system needs to scale with a response in terms of milliseconds to handle a few hundreds of trillions of requests per day [3]. Furthermore, distributing the computing on a Cloud platform [12] offers several benefits over the existing computing models such as dynamic scalability [13], rapid elasticity and pay per usage [14] assistance. It provides the users with an illusion of infinite storage and computing resources. Computing resources such as memory, CPU, servers and platforms can be used in a utility like manner [15] by paying only for the amount of usage of these resources. By utilizing Cloud as the computational model for biometric signature verification we would be able to address several additional issues like cost and energy optimization along with the major issue of huge information management. The information about each individual can be easily offloaded to a third party Cloud and managed efficiently.

Thus, this paper makes use of Cloud and parallel computing frameworks such as Apache Hadoop(for development of the proposed model) which is gaining popularity as a technology for handling large volumes of heterogeneous data [16]. It discusses about the existing approaches to mobile biometric authentication and then presents BAMCloud, a novel Cloud based mobile biometric authentication framework. BAMCloud provides a scalable and efficient approach for biometric signature verification. It involves data capture through handheld devices and then this data is offloaded to a third party Cloud where the data preprocessing and training is also performed. Finally, the data is queried for using Apache Hive [17]. The efficiency of the framework lies in the performance improvements gained over the existing ones in literature. A comparative cost and benefit analysis has been presented for the system. However, there are certain limitations while include data upload as a bottleneck in our system. Following are the contributions of this paper.

TABLE I
COMPARISON OF DYNAMIC V/S NON DYNAMIC SIGNATURE VERIFICATION

| Dynamic signature verification | Non dynamic signature verification |
| --- | --- |
| It is based on behavioural biometric | It is based on structural biometric |
| Features are extracted from the shape, speed, pressure, time taken by the person and his pen | Features are extracted only from the shape of the signature |
| It requires special digital surface for example digitizing tablet and pen | It does not require any special hardware |
| It is comparatively harder to forge | It is comparatively easier to forge |
| Example of features are x, y coordinates, pen pressure, time taken to sign, acceleration of the pen, pen tilt angles etc. | Example of features are the different components that can be extracted through an image like histogram, x, y coordinate etc. |
| Accuracy is around 99% [2] | Accuracy is around 95% [18] |

- Proposal of BAMCloud for meeting the data storage and processing requirements of data shoot-up in biometric signature samples.
- Development of an efficient parallel algorithm for preprocessing of biometric samples. A speedup of 8.5x was achieved which is better than the sequential algorithms existing in literature.
- Validation of BAMCloud through rigorous experimentations, which revealed that BAMCloud performs better than the existing systems (EER achieved was 0.24, which is lower than the existing systems).
- Cost benefit analysis of BAMCloud which shows that it requires lesser implementation and running cost as compared to the existing biometric approaches.

The rest of the paper is organized as follows Section 2 provides a summary of the existing approaches to mobile biometric authentication. Section 3 discusses the preliminary assumptions, notations and modeling used. Section 4 presents the proposed framework. Section 5 discusses the performance analysis along with experiments and implementation done. Furthermore, the cost and benefit of the system is also done in this section. Finally, the paper is concluded in section 6.

## II. LITERATURE SURVEY

Dynamic signatures are much more than static signatures in which just the shape of the signature is recorded. In dynamic signatures, the signatures are captured using touch sensitive digitizing tablet suitable for recording behavioral information such as pressure, velocity, pen-tilt angles, total time taken to sign and acceleration. Signatures captured in controlled environment with a pen tablet is relatively easier to implement but in touch enabled mobile devices, the quality of signature captured is not as good because the pen pressure and pen-tilt angle information is not present in mobile devices [2]. Table I shows a comparison between dynamic and non dynamic signature In the proposed approach we have used these features along with additional features like acceleration, magnetic field and angular velocity. Research carried out in the field of biometric signature verification on mobile devices is scarce. In general the signature data is captured using a pen tablet [19].

TABLE II
NEXT GENERATION REQUIREMENTS OF BIOMETRIC SYSTEM

| Requirements | Demands | Cloud Solutions |
|---|---|---|
| Data capture at different locations | Meet the biometric demands of constantly moving mobile society. | To capture data through handheld devices and store it on Cloud |
| Data acquisition through different devices | Overcome interoperability issues, Incorporate anti-spoofing techniques and Handle live data | Ability to perform pre-processing and handling of large live data |
| Minimization of biometric system response time | Increase need for agility of the system and accommodate constant need for authentication | Performs parallel and scalable distributed processing of signature samples |
| Platform independent biometric systems | Interoperability in authentication methods | Provision for platform as a service (PaaS) |
| Provision for data sharing amongst multiple applications | Sharing of data across different organizations and secured information exchange. | Secure sharing of information |
| Need for controlled parallelism | Use of techniques that are free from scaling bottlenecks and robust and failure free system | Availability of scalable infrastructure and ability to handle system failures |
| Quick response to user queries | Scalable query processing and parallel data analysis and query execution | Cloud techniques such as Hive and Pig, that can perform distributed query processing |

Although there are some works that involve capturing biometric signature through alternative devices [10], [1]. The conventional pen tablets usually capture more information than mobile devices; these devices capture information like pen pressure, pen trajectories, pen orientation, pen-ups and pen-downs along with others. Different algorithms have been applied for biometric signature authentication and verification purpose [20], [21]. Dynamic Time Warping (DTW), Hidden Markov Model (HMM), Artificial Neural Network (ANN), Support Vector Machine (SVM) are the mostly used algorithms in the literature [21]. The effects of using different signing space has been studied but not in context of hand held devices [22]. For the proposed work, different hand held devices like smart phones, PDAs and tablets were used to capture the biometric signature data. We have also captured orientation angles using the sensors present in almost every mobile device to achieve a better accuracy value compared to the work given in [10].

Next generation of biometric systems need to adapt themselves to the latest trends and technologies and cater to the needs of growing rate of data. Some of the next generation requirements of a biometric system along with solutions are listed in table II. Therefore, we need to develop systems that are interoperable with the existing biometric systems which are currently being used in forensics, industrial organizations, banking sectors and academics. These systems need to be flexible, expandable, scalable and highly dynamic to accommodate increasing demands of biometric technology and standards. Although the existing systems have used biometric techniques but most of them lacked these features and were not able to meet these requirements.

Hence, we can leverage the advantages and features offered by Cloud computing to tackle these requirements.

Virtualization is the key concept behind Cloud computing [23], [24] as it offers various features such as infinite data storage and memory along with reduced costs of dedicated servers. There are also several problems with adopting Cloud computing [25] for biometric data which pose as threat to data such as data privacy, VMware escape and mobility issues. Confidentiality of biometric data in Cloud can be enhanced using biometric encryption [26]. The authors in [1] used Amazon Cloud for performing biometric identification on different sized data and achieved a cost optimal solution for it. A cancellable biometric authentication approach [2] can be used for management of biometric data. In this approach a distorted biometric image can be used for data authentication [27] and data hiding is performed for embedding demographic information in biometric samples.

Hadoop is a popular programming framework becoming synonymous with Cloud platforms. Authors in [28] have used Hadoop MapReduce to implement iris biometric system. The results show increased speedup and efficiency over the sequential approaches. Authors in [29] have also used Hadoop for processing of biometric data at large scale. They have leveraged the use of public Cloud services provided by Amazon EC2 for validating their approach in order to achieve enhanced performance. Therefore, it can be concluded that most of the work done in literature for handling large scale biometric data using Hadoop is done on systems that are not socially acceptable and requires expensive capturing devices.

All-Pairs is an abstraction developed by authors in [6] which makes use of data intensive computing performed on campus grid. They have implemented their approach on biometric face samples and it was deduced from their findings that biometric face comparison function took one second to compare two 1.25MB images. Authors in [7] have worked on animal identification using Cloud technology in order to handle large scale collaborative wildlife monitoring with citizen scientists. Authors in [9] have used an amalgamation of Cloud and biometric fingerprint recognition using assembled geometric moment and Zernike moment features for secure communication in Cloud computing. After systematic study of these approaches [6], [7], and [9], it was deduced that these approaches were expensive and the BAMCloud lead to a significant amount of reduction in costs incurred for implementing biometric systems.

Thus, literature has laid its hands on topics like data identification and security of biometric data in Cloud. Moreover, most of the work done is based on static biometric samples; this paper focuses on user authentication aspect of mobile biometric through the use of Cloud techniques. However, issues concerning the security of biometric data are beyond the scope of this work. The work done in this paper uses artificial neural network as a model to train the biometric signature data spread over a Hadoop cluster and it uses signature samples for authentication purpose.

## III. PRELIMINARY CONCEPT

### A. Assumptions

For the implementation of proposed framework we consider an environment consisting of a set of handheld devices at various positions spanning across an area of 15000 km₂. It is assumed that all these devices are connected via the internet at all times. There is a set of D devices that are accessible to U users. Thus, every entity in our biometric system consist of a tuple T where T = {U, D}. Therefore, every tuple T belonging to {T₁, T₂,...} is connected to a high performance computing Cloud (HPC) cluster. It is assumed that as soon as the signature is captured it is transferred to a Cloud and its training and processing is offloaded to another private Cloud cluster. Time delay and network latency is ignored in our system and this duration between data capture and transfer is almost negligible.

The signatures have been captured in two different sessions from each user in a difference of 5 days to incorporate the intra class variations that might occur due to the mood and emotional imbalance. It is also notified that the data has been captured statically i.e. the subject was stationery at the time of collection of samples and the data is collected from 100 fixed locations. The notations used in this paper have been described in table III.

### B. Data Model

BAMCloud operates over a set of entities where each entity is a representative of real world objects.

#### 1) Signature Sample modeling:

Each representative sample collected in this framework is composed of dynamic parameters representing acceleration ($\alpha$), magnetic field ($\mu$), orientation ($\Phi$) and angular velocity (V). Thus, each sample is a quadruple i.e. Sample $S_{ij} = \{\alpha, \mu, \Phi, V\}$, where $S_{ij}$ is the $j^{th}$ sample of $i_{th}$ user, $0 < j \leq 40$ and i can be any large value since our framework is dynamically scalable giving users an illusion of infinite storage capacity. Each element of this quadruple is further divided into triplets.

TABLE III
FEATURES RECORDED WHILE SIGNING

| Parameters | Features | Unit |
|---|---|---|
| Acceleration | $X_\alpha$ | m/s2 |
|  | $Y_\alpha$ | m/s2 |
|  | $Z_\alpha$ | m/s2 |
| Magnetic Field | $X_\mu$ | Mt |
|  | $Y_\mu$ | Mt |
|  | $Z_\mu$ | Mt |
| Orientation | Azimuth | Degrees |
|  | Pitch | Degrees |
|  | Roll | Degrees |
| Angular Velocity | $X_v$ | rad/s |
|  | $Y_v$ | rad/s |
|  | $Z_v$ | rad/s |

Thus, $\alpha$ is a set consisting of triplets where $\{X_\alpha, Y_\alpha, Z_\alpha\} \in \alpha$, $\mu$ is also a triplet where $\{X_\mu, Y_\mu, Z_\mu\} \in \mu$, $\Phi$ is composed of three elements as well {Azimuth, Pitch, Roll} $\in \Phi$ and angular velocity V represented by $\{X_v, Y_v, Z_v\}$ measured in radians per second.

#### 2) Feature Extraction Model:

The feature extraction model adopted in this approach consists of a distributed approach where the pre-processing of data is done in a distributed manner.

**Problem Definition of data sample:**

*Given 'm' samples, with n features representing N dimensional space of data, the problem lies in how to project this data to a lower dimensional space for a very large value of n while preserving the similarities and variations in the data sample.*

The solution to this problem lies in adoption of a mechanism which preserves the behavior of data while adhering to the costs required for storage and analysis and one such mechanism prevalent in the literature is Principle Component analysis (PCA).

## IV. BAMCLOUD

To overcome the data management and performance issues arising from the implementation of biometric system in hand held devices, we propose BAMCloud. It takes advantage of infinite storage and computing capacity offered by Cloud technology. The storage and processing of dynamic signature samples takes place at a third party location provided by the Cloud service provider. Fig. 1 provides an overview of BAMCloud. The framework is divided into five phases: (1) data capture, (2) data storage, (3) data pre-processing, (4) training, testing and storage of the model and (5) query phase.

### A. Phase 1 Data Capture

In this phase data is captured through touch sensitive hand held devices, and all the relevant features are acquired via the sensors associated with the device. The total number of users in our experiment is 630. Each of these users was asked to give 20 genuine signatures, and other users were asked to produce skilled forgery for the user, 20 forged signature samples were captured. Therefore, a total of 40 signature samples (20 genuine and 20 forged) were captured for each user. The captured features are listed in table III.

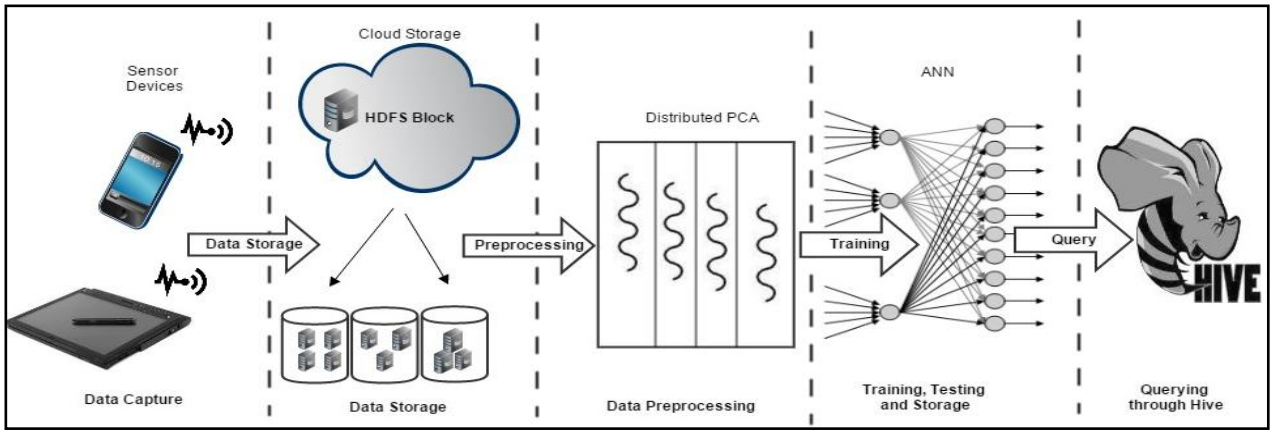

Fig. 1: BAMCloud: Cloud based mobile biometric authentication framework

### B. Phase 2 Data Storage

After the data has been captured, it is then stored in the HDFS Cloud [30]. HDFS is a distributed file system which can be deployed on low cost commodity hardware. It is highly fault tolerant with high throughput and has the ability to store very large datasets. Since the data of signature samples collected by us is very large therefore, HDFS comes up as a viable solution. It easily allows scaling upto hundreds of nodes in a single cluster thus supports up to gigabytes or terabytes of data [31].

### C. Phase 3 preprocessing

This phase performs pre-processing of signature samples that were collected from mobile devices. In order to carry out pre processing of these samples two techniques were adopted. PCA has been used in both the approaches i.e. sequential and distributed. The reason why PCA has been used is to take advantage of time and storage space reduction that it offers. Since, we are using neural network to perform modeling of the system therefore, PCA is a feasible option because of its multi-co linearity removal feature which aids in enhancing performance of machine learning model.

In order to analyze and compare the performance gain obtained by using the distributed approach, firstly, data was pre-processed using algorithm 1 which adopted the use of sequential PCA. Secondly, the same data was pre-processed in a distributed manner via algorithm 2.

**Definition 1:** *Let Getsigsamples represent the method for capturing the dynamic and behavioral features of the signature of mobile users.*

In algorithm 1 called as ALGOSigPreprocess, in steps 2-6, M signature samples are collected for every individual using mobile device. There are total of N users against which the samples are obtained. Thus, there are M x N datasets collectively and $I_d$ is the input data which is collected. After collection of datasets covariance is computed which is represented by $\mathcal{C}_{ev}$ and then square root of diagonal elements of covariance matrix is obtained for the captured data represented by S and S' is the transpose of the square matrix.. After this correlation i.e. $\mathcal{C}_{er}$ is found out using step 9 which is further used to find out the PCA [32] ($\mathcal{P}_{caccef}$) of the data through singular value decomposition (svd).

Algorithm 2 called as ALGODistSigPreprocess is a MapReduce based distributed version of algorithm1. In steps 2 to 6 data is captured in a manner similar to the one in algorithm 1. After that covariance MapReduce function is

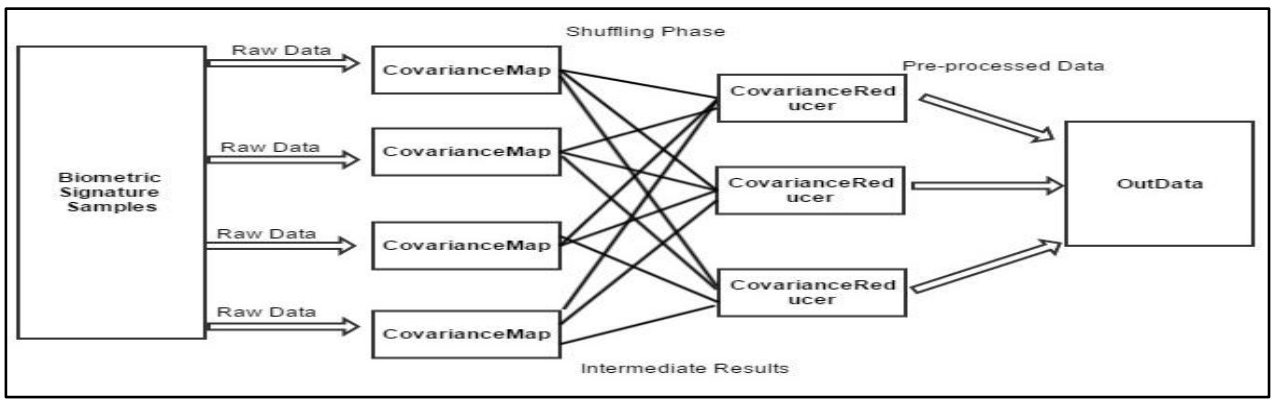

Fig. 2: MapReduce Modeling of ALGODistSigPreprocess

---

**ALGORITHM 1**
**ALGOSigPreprocess**

---

**Input:** $N$: number of users, $M$: number of signature samples of each user
**Output:** $pcacoef$ : matrix of preprocessed samples
1.  **Begin:**
2.      **For** i=1 to $N$
3.          **For** j=1 to $M$
4.              Getsigsamples($Sdij$)/*captures dynamic features of a signature of an individual, $Sdij$ is the $j^{th}$ sample of $i^{th}$ user*/
5.          **End** For
6.      **End** For
7.      $Cov$=covariance($Sd$) /*calculate covariance of the input matrix*/
8.      s=$sqrt$(diagonal($Cov$)) /*calculating the square root of the diagonal elements of the covariance matrix*/
9.      $Cor$=$Cov$/s*s' /* calculating correlation from covariance matrix and product of square root matrix obtained from step 8 and its transpose*/
10.     $Pcacoef$=$svd$($Cor$) /*performing PCA using singular value decomposition on the correlation matrix obtained from step 9*/
11.     **Return** $Pcacoef$
12. **End**

---

**ALGORITHM 2**
**ALGODistSigPreprocess**

---

**Input:** N: number of users, M: number of signature samples of each user
**Output:** $pcacoef$ matrix of preprocessed samples
1.  **Begin**:
2.      **ParFor** i=1 to N
3.          **ParFor** j=1 to M
4.              Getsigsamples($I_{dij}$))/*captures dynamic features of a signature of an individual, $Sdij$ is the $j^{th}$ sample of $i^{th}$ user in a distributed manner*/
5.          **End ParFor**
6.      **End ParFor**
7.      $Od$ = mapreduce ($Sd$; covariancemapper, covariancereducer) /* data output after running mapreduce job on input data samples*/
8.      $Cov$=covariance($Od$) /*calculate covariance of the input matrix*/
9.      s=$sqrt$(diagonal($Cov$))/*calculating the square root of the diagonal elements of the covariance matrix*/
10.     $Cor$=$Cov$/s*s' /* calculating correlation from covariance matrix and product of square root matrix obtained from step 9 and its transpose*/
11.     $Pcacoef$← $svd$($Cor$)) /*performing PCA using singular value decomposition on the correlation matrix obtained from step 10*/
12.     **Return** $Pcacoef$
13. **End**

---

called.

MapReduce modeling of the proposed approach given by Step 7 is elaborated in Fig. 2. According to this figure biometric signature sample data is passed through a MapReduce programming model. This MapReduce programming model is used for parallelizing the processing of our data since the data is of huge size. It uses key value pair as a data type in general. MapReduce programming of our proposed approach is divided into two functions: a CovarianceMapper function and a CovarianceReducer function.

The job of the MapReduce framework is to execute these functions in parallel on different machines, the number of machines depend upon the size of data being executed. The exquisiteness of this approach lies in the fact that MapReduce is highly scalable. The data is firstly processed in parallel by the CovarianceMapper and then this output is recombined by the CovarianceReducer function. After completion of the MapReduce phases, PCA coefficients are computed at different nodes and the consolidated result is returned by step 12.

### D. Phase4 Training, Testing and Storage

During this phase the pre-processed data is trained using feed-forward backpropagation neural network [33], then the network is tested using stratified subset of the data set. Finally, these trained networks with acceptable FAR/FRR rates are stored back on HDFS.

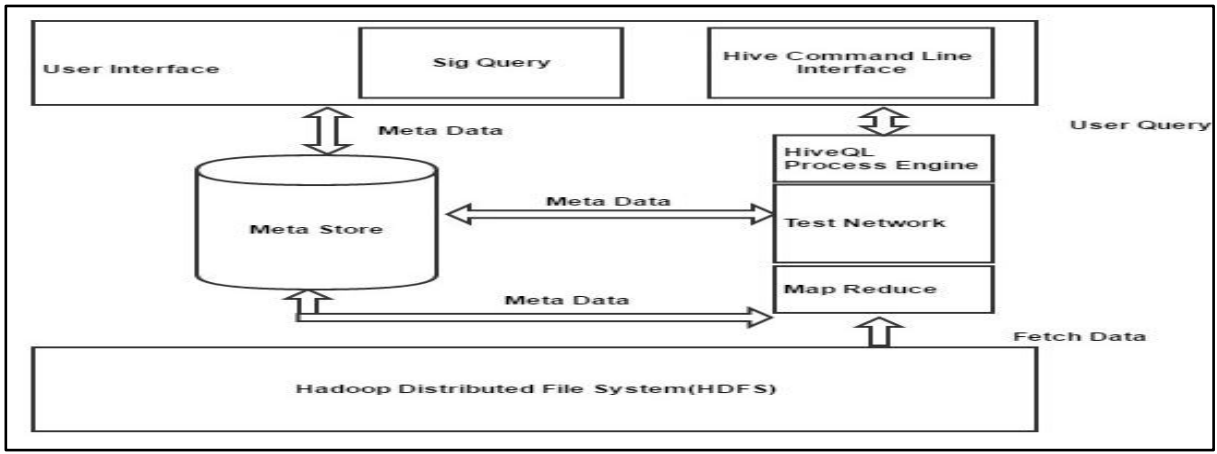

Fig. 3: Hive Architecture for SigQuery

**ALGORITHM 3**
**ALGOTrainSample**

**Input**: $\mathcal{P}_{caccef}$: matrix of preprocessed samples obtained from phase 2
**Target**: target matrix
**Output**: Net1: trained network
1.   **Begin:**
2.      **Training:**
3.        **For** i=1 to N
4.          **For** j=1 to M
5.            Input1$_{ij}$← $\mathcal{P}_{caccef_{ij}}$ /* inputting preprocessed signature samples obtained from phase 2*/
6.          **End For**
7.        **End For**
8.          net1=ffn$_n$() /*network creation using backpropagation algorithm on a feedforward neural network*/
9.          Net1=train(net1,Input1,target) /*training is performed on the created network using input and target vectors*/
10.      **End** of training
11.      **Return** Net1 /* trained network is returned as output*/
12.   **End**

**Definition 2:** *ffnx is a function that implements backpropagation algorithm on a feed forward network with input layer of i nodes, n hidden layers with h nodes and an output layer of two nodes representing genuine or forged.*
Definition 3: fdfnx performs backpropagation algorithm on a distributed feedforward network, the distribution is decided based on the size of the dataset.

Algorithm 3 called as ALGOTrainSample has been used to do the training on the dataset. Input1 denotes the network input to be trained. In steps 2-7 the preprocessed data corresponding to the jth sample of ith user has been taken as input in Input1ij. This input data is to be treated as the training set. Function for feed forward neural network represented by ffnx is then applied along with backpropagation algorithm for training purpose as shown in step 8. In step 9 training is performed using input and target vectors and finally the trained network is returned.

Algorithm 4 ALGODistTrainSample is the distributed version of algorithm 3. In this algorithm, the training is done on the datasets in a parallel manner. For training of the proposed feedforward neural network based system, different learning algorithms for example Levenberg-Marquardt, Conjugate gradient backpropagation, Bayesian regularization, resilient Backpropagation and Bayesian regularization were applied in order to recognize pre-processed biometric signature data and the trainlm learning algorithm resulted in the best results in the form of EER. All the processing has been done on Hadoop. This algorithm takes the preprocessed signature data i.e. Pcacoefij as input. Function for distributed feed forward neural network represented by fdfnx is then applied along with backpropagation algorithm for training purpose as shown in step 10. The output of this data produces

**ALGORITHM 4**
**ALGODistTrainSample**

**Input**: $\mathcal{P}_{caccef}$: matrix of preprocessed samples obtained from phase 2
**Target**: target matrix
**Output**: $\mathcal{T}_{Global}$: trained network
1.   **Begin:**
2.      **Parallelized training:**
3.        **For** i=1 to N
4.          **For** j=1 to M
5.            Input1$_{ij}$← $\mathcal{P}_{caccef_{ij}}$ /* inputting preprocessed signature samples obtained from phase 2 in a distributed manner*/
6.          **End For**
7.        **End For**
8.          $\mathcal{T}_{Global}$= Φ /* initially the data set for trained network is empty*/
9.        **For** 1 ∈ L do /*repeat for all local networks*/
10.          net1=fdfnx() /*network creation using backpropagation algorithm on a distributed feedforward neural network*/
11.          $\mathcal{T}_s$=train(net1,input,target) /*training is performed on the created local networks using input and target vectors*/
12.          $\mathcal{T}_{Global}$ ← $\mathcal{T}_s$∪ $\mathcal{T}_{Global}$ /*global network is created by combining all the local networks*/
13.        **End of parallelized training**
14.      **Return** $\mathcal{T}_{Global}$ /* trained global network is returned as output*/
15.   **End**

a global network ($\mathcal{T}_{Global}$) that is trained on the dataset. This global network is obtained by combining all the local networks i.e. $\mathcal{T}_s$ and the trained network is then returned in step 14.
The following section briefly describes the training algorithms that were applied.

*1) Levenberg-Marquardt backpropagation*

Levenberg-Marquardt backpropagation algorithm is used to achieve second-order training pace without computing the Hessian matrix. The Hessian matrix can be represented by equation 1:

$$H = J^T \bullet J \qquad (1)$$

Where J is the Jacobian matrix [26].
And its gradient is computed as given in equation 2:

$$g = J^T \bullet e \qquad (2)$$

The Jacobian matrix contains the first order network error derivatives where 'e' denotes a set of errors. A back-propagation algorithm can be used to compute the Jacobian matrix which is easier than Hessian matrix computation. This algorithm uses the following update as shown in equation 3 to approximate the Hessian matrix:

$$x_{k+1} = x_k - [J^T J + \mu I]^{-1} J^T e \qquad (3)$$

*2) Conjugate gradient backpropagation*

In basic backpropagation algorithm, the weights are adjusted in the direction of negative gradients. It can be noted, that even if the performance decreases most rapidly in this direction but it does not necessarily converge that rapidly. Therefore, in Conjugate gradient backpropagation algorithm, the weights are adjusted in conjugate directions, and it takes less time to converge than steepest descent methods [26].

Neural Network needs to provide second order information to conjugate gradient methods, but it needs only O(N) memory where N is the number of weights in the neural network [1]. In most of the conjugate gradient methods searching is started out in the steepest descent direction at first and then search is continued in the conjugate direction to determine the step-size, and then the step size is adjusted in next iteration [26]. Here, N is the number of weights in the neural network [1].

In most of the conjugate gradient methods searching is started out in the steepest descent direction at first and then search is continued in the conjugate direction to determine the step-size, and then the step size is adjusted in next iteration [26].

### 3) Resilient backpropagation

The resilient backpropagation algorithm is also based on the conventional backpropagation algorithms that compute the errors of the network and tries to minimize it by modifying the weights of the network.

### 4) Bayesian regularization

In order to design a network that has minimum errors and that generalizes well, Bayesian method is proposed to constrain the network parameter size. The system will find difficulty in its implementation if the network is over-sized. With the inclusion of regularization the objective function can be represented by equation 4:

$$F = \gamma E_D + (1 - \gamma) E_w \qquad (4)$$

Where $E_w$ is the sum of squares of the parameters of the network and $\gamma$ is the ratio of the performance. This optimal parameter can be determined using Bayesian rule:

$$P(W \mid D, \gamma, M) = \frac{P(D \mid W, \gamma, M) P(W \mid \gamma, M)}{P(D \mid \gamma, M)} \qquad (5)$$

Where, M is the neural network model, D represents the data, W is the set of working parameters [2]. Bayesian regularization in neural network reduces the requirement of extended cross-validations. This algorithm is difficult to overtrain and overfit [16].

### 5) Gradient descent backpropagation

Gradient descent is a first-order optimization problem, In order to find the local minima of a function; step is taken towards the negative direction of the gradient at that point. gradient descent may or may not find global minima. Let f(x) is a minima function of a variable x, x is n-dimensional. Then $-\nabla f(x)$ represents the maximum descent direction. Thus, at a point x0, the gradient can be calculated using equation 6.

$$x_0 - t \nabla f(x_0) \qquad (6)$$

Let $t_1$ is the point at which function f has a minimum value. The value of x1 is calculated using equation 7.

$$x_1 = x_0 + t_1 \nabla f(x) \qquad (7)$$

Then the local minima of f(x) are close.

### E. Phase 5 Query

After the data is captured, pre-processed and trained, it is then stored on Cloud and queried for. The query is performed to check whether the user is a legitimate one or not. The process of querying is carried out by Hive as the data is stored on HDFS. Hive [17] is a data warehouse software that provides provision for querying of large datasets which resides on a distributed storage system. The data is queried into HDFS using HiveQL [17].

Fig. 3 shows the Hive architecture for the proposed SigQuery i.e. query process for signature data. The end user firstly issues a SigQuery at the Hive command line interface. The processing then goes to the HiveQL process engine where the query is processed which involves testing of the network through MapReduce programs. It also involves data retrieval from HDFS and meta store which contains meta data information. To validate a transaction, the user is asked for a signature, that he can input using his mobile device, and then this signature is preprocessed. After the preprocessing a SigQuery is issued at the Hive command line interface and then sent to the HiveQL process engine for testing purpose, where it is checked against the signatures of registered users stored in HDFS to find out whether the signature is genuine or forged.

## V. PERFORMANCE ANALYSIS

This section performs the analysis of the algorithms proposed in BAMCloud by comparing them with the ones existing in literature. It also discusses the experimental setup and evaluation metrics used.

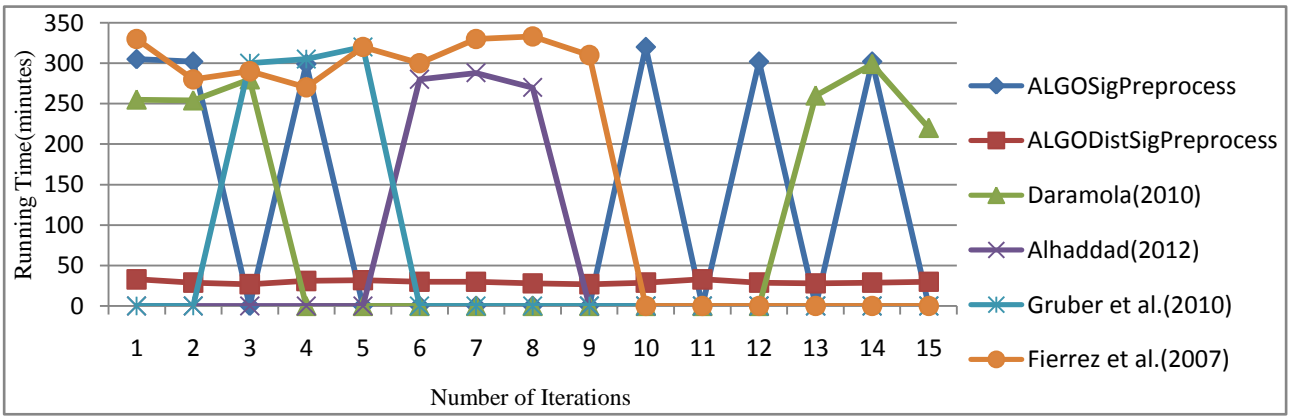

Fig. 4: Plot of running time of ALGOSigPreprocess versus other algorithms in literature. Time zero means that system crashed when an algorithm was executed

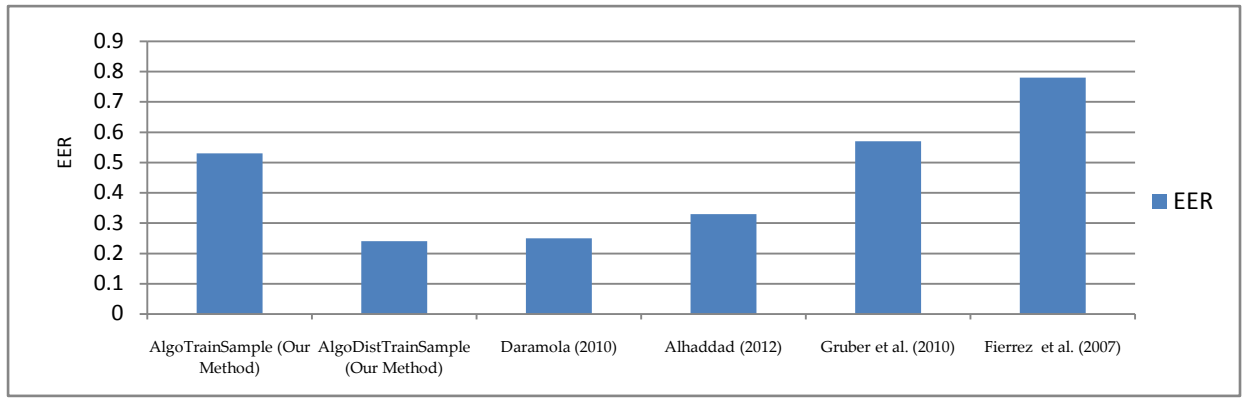

Fig. 5: Performance Evaluation of BAMCloud versus Baseline biometric signature methods in terms of EER

## A. Experimental Setup

In this section, the experimental setup of all the phases is explained.

### 1) Baseline System

The implementation of BAMCloud was done on a 16 cabinet cluster based on commodity off shelf building blocks. It has a total of 397 nodes using Haswell processors which are well suited for HPC requirements. The number of cores per node was 24 and memory available per node was 128 GB. Hence, the total number of cores available in the cluster was 9528 (397 x 24) and total memory available was 50816 (397 x 128) GB. The experimental data sets used by us are real handwriting data collected on a mobile device.

### 2) Data capture and dataset used

Mobile devices have been used, working on different platforms such as iOS, android and windows to capture signature samples. The devices used were handheld and belonged to four different well known brands. The brands and their respective models were: Xiomi's Mi 4i, Samsung Galaxy s3 and Note 2, Apple's iPad and iPhone 4s and Motorola's MotoG. The users were asked to use both stylus and fingertips to record their signatures. Signature samples from 630 users were collected, each user was asked to give 20 genuine signature and other users were requested to give 20 forged signatures for the user. Therefore, 40 signature samples were stored against each user thus total number of signature

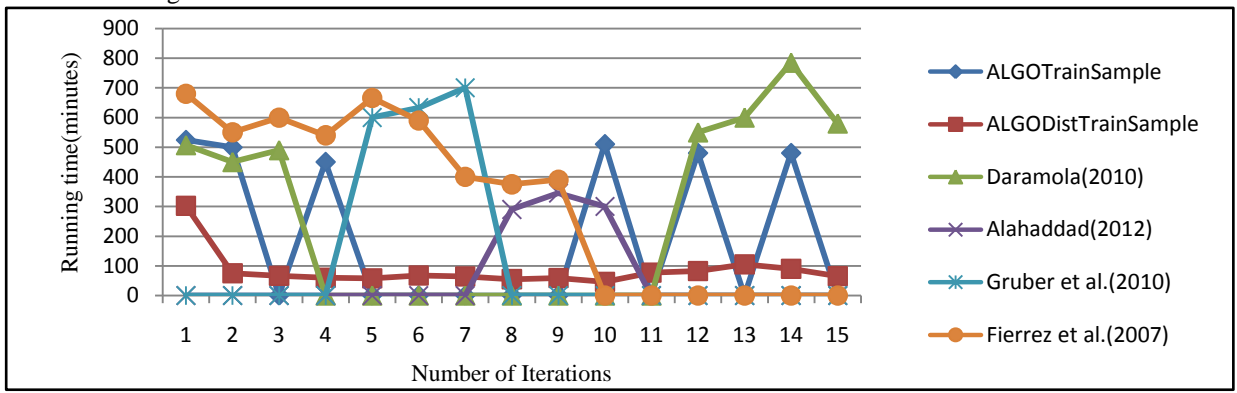

Fig. 6: Plot of running time of ALGODistTrainSample versus other algorithms in literature. Time zero means that system crashed when an algorithm was executed

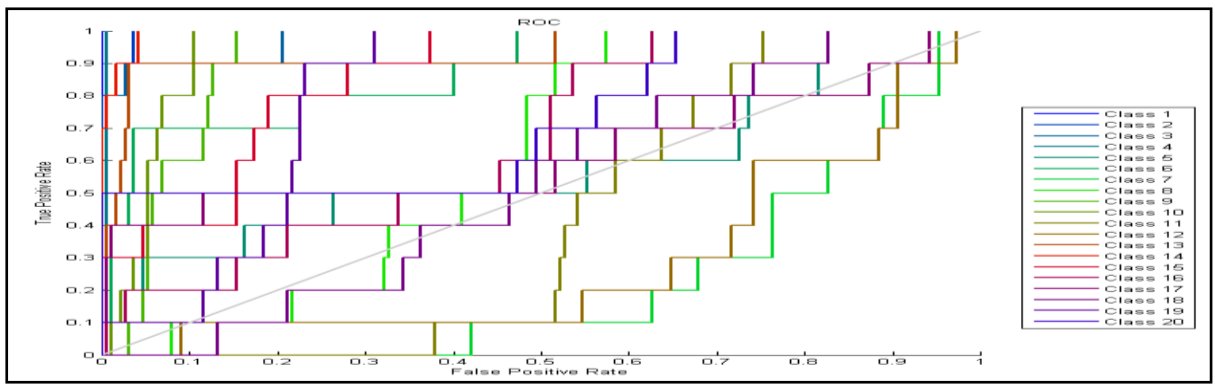

Fig. 7: ROC curve of first 20 users

TABLE IV
COMPARATIVE ANALYSIS OF PERFORMANCE OF BAMCLOUD

| Title | Technology/ Platform used | Performance (%) | References |
|---|---|---|---|
| Online Signature Verification in Banking Application: Biometrics SaaS Implementation | Microsoft Azure | 94.25 | [35] |
| Signature Verification SaaS Implementation on Microsoft Azure Cloud | Microsoft Azure | 94.25 | [36] |
| Multi-Factor Authentication on Cloud | Multi-factor key (handwritten signature + standard knowledge factor) | 98.4 | [37] |
| BAMCloud (Our method) | Amazon EC2 | 96.23 | |

samples stored was 252000.

### 3) Feature Extraction

Feature extraction was performed using distributed PCA on a Hadoop distributed file system (HDFS). The data was distributed on a high performance computing cluster with 397 nodes. The processing was carried out in Hadoop environment using MapReduce programming model. The advantage of using this environment is that computations can be performed on commodity hardware thereby saving on the inventory cost. Apart from this it offers high reliability, scalability and flexibility. Another advantage of this method is that computations can be performed fast, since there is no data movement overhead as Hadoop supports bringing computation to data rather than data to computation.

### B. Experiments and Implementation

#### 1) Evaluation Metrics

The following metrics have been adopted in order to check the authentication and performance of the system.

1. *False acceptance rate (FAR)*
   FAR measures the accuracy of the biometric system by providing the probability of the cases when the system accepts an incorrect output (forged sample), FAR can be represented using the terms false positives (FP) and true negatives (TN) as given in equation 8.

$$FAR = \frac{FP}{FP + TN} \quad (8)$$

2. *False Rejection rate (FRR)*
   FRR measures the accuracy of the biometric system by providing the probability of the cases when the system rejects a correct output (genuine sample), FRR can be represented using the terms false negatives (FN) and true positives (TP) as given in equation 9.

$$FRR = \frac{FN}{FN + TP} \quad (9)$$

3. *Equal Error Rate (EER)*
   The value when both FAR and FRR becomes equal is the EER of the system. It can be represented using equation 10.

$$EER = FAR|_{FAR=FRR} = FRR|_{FRR=FAR} \quad (10)$$

4. *Speedup*
   The speedup achieved is defined as the ratio of running time of sample on a single processor to that of running time of sample on 'n' nodes. The speedup S (n) is defined where, T(S) is running time of a single processor and T (n) is the running time of the system on 'n' nodes.

5. *Signature Feature Variation*
   This metric measures the effect of varying signature features in terms of equal error rate (EER). It involves altering the features and analyzing their impact on the performance of the system. The features include acceleration, magnetic field, orientation and angular velocity

6. *Cost*
   This metric analysis the effectiveness of system in terms of cost incurred for running and implementation of the system. The cost is measured in terms of USD.

### C. Results

#### 1) Data pre-processing

In order to perform data pre-processing PCA has been used. There were two approaches adopted, Firstly, ALGOSigPreprocess was used which is a sequential version of PCA. It was observed that the system crashed more than 50 percent of the times when this algorithm was run. This can be attributed to the incapability of the system to handle the data of such huge volume. Therefore, to overcome this limitation ALGODistSigPreprocess which is a distributed version of ALGOSigPreprocess was adopted. The system on running ALGODistSigPreprocess was able to pre-process the entire data set by spanning across 28 nodes. The average time taken to perform PCA using ALGODistSigPreprocess was 29.6 minutes. In order to further validate the applicability of the proposed approach a plot of running time achieved using the ALGODistSigPreprocess versus other algorithms in literature was plotted as shown in Fig. 4.

TABLE V
SUMMARY OF SPEEDUPS

| Algorithm | Speedup |
|---|---|
| ALGODistSigPreprocess | 10x |
| AlgoDistTrainSample | 7x |
| BAMCloud | 8.5x |

The experiments were conducted repeatedly 15 times and it was observed that the system could achieve a speedup of 10x using algorithm 2.Therefore, at the end of pre-processing, the data was reduced to 25 percent (approximately) of the original size thereby reducing the storage space. The reduction in storage space leads to faster processing and cost optimizations.

### 2) Training and Testing

After the data is pre-processed and brought in a uniform format, the data was trained using Levenberg-Marquardt Backpropagation algorithm on a feed forward network. Fig. 5 shows the results of the two proposed training algorithms, AlgoTrainSample and AlgoDistTrainSample. From the results it can be concluded that AlgoTrainSample has an EER of 0.53 and it performs better than methods used in [38] and [39] but cannot perform better than [40] and [41], so a distributed approach is adopted and the performance with an EER of 0.24 was achieved which was the best performance achieved by the system.

Fig. 6 shows plot of running time of AlgoDistTrainSample versus other algorithms in literature and it was observed that the speedup achieved by our distributed training algorithm using equation 8 was 7x (approximately) as compared to other algorithms in literature.

Fig. 7 shows the ROC curve of first twenty user data, after applying AlgoDistTrainSample to it. ROC curve represents the tradeoff between the false acceptance rate and the false rejection rate. For the shown data of 20 users we get good curves for 17 users and for the other 3 users it was observed that the false positive rate is high. In Fig. 8, the ROC curves of both the algorithms are given. The ROC curve of AlgoDistTrainSample in Fig. 8b is better than that of AlgoTrainSample given in Fig. 8a. It can be seen that when artificial neural network was used on a standalone machine, the performance of the system is comparable with the best methods in the literature. In AlgoDistTrainSample we use the features of Hadoop to distribute the training samples and thereby the performance of the biometric system improves by 29 percent.

Table V, shows the overall speedup achieved by BAMCloud which is the average of speedups of ALGODistSigPreprocess and AlgoDistTrainSample. Thus, the speedup achieved by the BAMCloud was 8.5 times faster than the other sequential approaches for biometric systems existing in literature. Hence, it can be concluded that using this approach, the ever growing data for biometric authentication system can be processed with a speed which is 8.5 times faster than the conventional systems. Also, during experimentations it was observed that most of the conventional systems failed when such heavy workload was processed.

### 3) Effect of varying the Signature features on the system

In this section the effects and impact of different features on AlgoTrainsample and AlgoDistTrainSample have been discussed, by altering the features used for classification purpose. From table VI, it can be seen that, features play a crucial role in the classification of signature samples, it is essential to analyze that which feature is more important and which one is less important for making the authentication system computationally inexpensive. We can see that the error rate increases as we exclude features like $X_\alpha$ and $Y_\alpha$, whereas features like $Y_v$ and $Z_v$ don't bring any change in the error rate of the whole system; hence it can be concluded that $Y_v$ and $Z_v$ features can be excluded from the database without effecting performance of the system. Apart from the difference in error rate, the table also demonstrates that the error rate achieved in distributed version of the algorithm is lower than that of the sequential one. Hence, AlgoDistTrainSample performs better than AlgoTrainsample even with varying features.

### 4) Cost and Benefit Analysis

Since the proposed system uses Cloud techniques and one of the major benefits associated with this technique is the cost apart from scalability and elasticity. Therefore, this section performs the cost and benefit analysis of the proposed system. The cost benefit analysis performed by us considers the hardware cost and the cost for processing the system. The analysis done takes into consideration two metrics. First metrics used is the cost comparison metrics, where a cost comparison is done for different biometric domains. Second metrics is the size of data being processed.

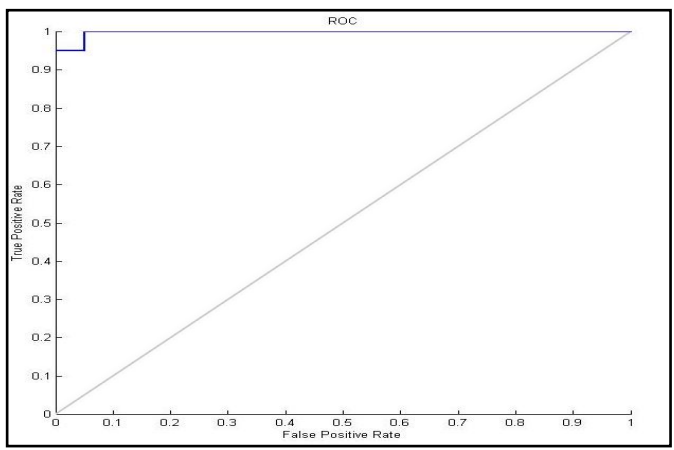

Fig. 8a

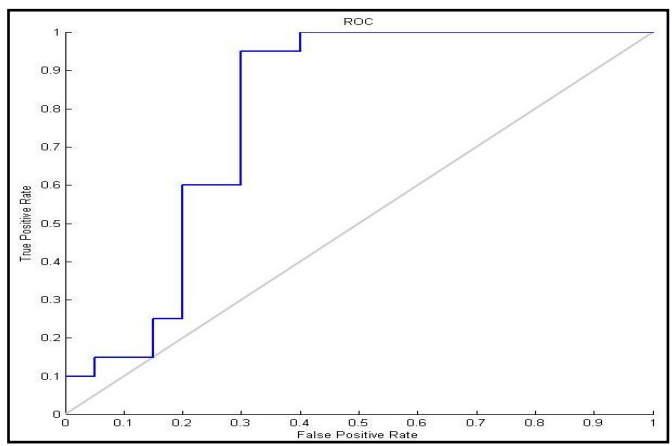

Fig. 8b

Fig. 8: ROC curves of the (a) AlgoTrainsample and (b) AlgoDistTrainSample

TABLE VI
EFFECT OF VARYING FEATURES ON SIGNATURE SAMPLE

| | Combination | $X_a$ | $Y_a$ | $Z_a$ | $X_\mu$ | $Y_\mu$ | $Z_\mu$ | Azimuth | Pitch | Roll | $X_v$ | $Y_v$ | $Z_v$ | EER |
|---|---|---|---|---|---|---|---|---|---|---|---|---|---|---|
| | 1 | √ | √ | √ | √ | √ | √ | √ | √ | √ | √ | √ | x | 0.57 |
| | 2 | √ | √ | √ | √ | √ | √ | √ | √ | √ | √ | x | √ | 0.58 |
| | 3 | √ | √ | √ | √ | √ | √ | √ | √ | √ | x | √ | √ | 0.57 |
| | 4 | √ | √ | √ | √ | √ | √ | √ | √ | x | √ | √ | √ | 0.59 |
| AlgoTrainSample | 5 | √ | √ | √ | √ | √ | √ | √ | x | √ | √ | √ | √ | 0.50 |
| | 6 | √ | √ | √ | √ | √ | √ | x | √ | √ | √ | √ | √ | 0.53 |
| | 7 | √ | √ | √ | √ | √ | x | √ | √ | √ | √ | √ | √ | 0.54 |
| | 8 | √ | √ | √ | √ | x | √ | √ | √ | √ | √ | √ | √ | 0.55 |
| | 9 | √ | √ | √ | x | √ | √ | √ | √ | √ | √ | √ | √ | 0.55 |
| | 10 | √ | √ | x | √ | √ | √ | √ | √ | √ | √ | √ | √ | 0.53 |
| | 11 | √ | x | √ | √ | √ | √ | √ | √ | √ | √ | √ | √ | 0.56 |
| | 12 | x | √ | √ | √ | √ | √ | √ | √ | √ | √ | √ | √ | 0.56 |
| | 1 | √ | √ | √ | √ | √ | √ | √ | √ | √ | √ | √ | x | 0.20 |
| | 2 | √ | √ | √ | √ | √ | √ | √ | √ | √ | √ | x | √ | 0.20 |
| | 3 | √ | √ | √ | √ | √ | √ | √ | √ | √ | x | √ | √ | 0.25 |
| | 4 | √ | √ | √ | √ | √ | √ | √ | √ | x | √ | √ | √ | 0.22 |
| AlgoDistTrain Sample | 5 | √ | √ | √ | √ | √ | √ | √ | x | √ | √ | √ | √ | 0.23 |
| | 6 | √ | √ | √ | √ | √ | √ | x | √ | √ | √ | √ | √ | 0.23 |
| | 7 | √ | √ | √ | √ | √ | x | √ | √ | √ | √ | √ | √ | 0.22 |
| | 8 | √ | √ | √ | √ | x | √ | √ | √ | √ | √ | √ | √ | 0.24 |
| | 9 | √ | √ | √ | x | √ | √ | √ | √ | √ | √ | √ | √ | 0.24 |
| | 10 | √ | √ | x | √ | √ | √ | √ | √ | √ | √ | √ | √ | 0.25 |
| | 11 | √ | x | √ | √ | √ | √ | √ | √ | √ | √ | √ | √ | 0.26 |
| | 12 | x | √ | √ | √ | √ | √ | √ | √ | √ | √ | √ | √ | 0.26 |

As the size of data increases so does processing requirement. Amazon Elastic Compute Cloud (Amazon EC2) is the Cloud computing service which has been taken for the purpose of analysis. It is a Cloud based web service which provides web scale computing capacity in a resizable manner. It allows its users to increase and decrease their compute capacity based on usage demand and pay on an hourly basis. For performing the cost and benefit analysis of our system, a metric termed as Total Cost (CostT) has been used, which is the total cost incurred for the implementation of the system and is given by the equation 12.

$$Cost_T = Cost_I + Cost_C \qquad (12)$$

Where $Cost_I$ is the cost of hardware for capturing data samples. $Cost_C$ is the total amount spent in performing the processing and running the virtual machines (VMs) on the Cloud providers' site. The analysis considers only the amount charged for running the VMs, other charges such as network and storage have been ignored as they depend on data requirement and applications communication. It should be noted that the users are charged only when the VMs start processing, no costs are applicable during booting of the machines. $Cost_C$ is further described by equation 13.

$$Cost_C = n_v \times c_v \times t \qquad (13)$$

Where, $n_v$ is the number of VMs required by the system $c_v$ is the cost of using a particular instance of VM depending on the usage. For example Amazon EC2 provides different types of instances such as T2, M3, C4, C3, R3, G2, I2 and D2. t denotes the time in hours for which a particular instance type was utilized. Equation 14 further describes t as:

$$t = t_c - t_s \qquad (14)$$

Where, $t_c$ is the time of completion and $t_s$ is the time of submission of the job.

Since our system makes use of commodity hardware and mobile phones. Therefore, the $Cost_I$ is almost negligible for our system and $Cost_C$ amounts to approximately 5000 USD. Fig. 9 shows the cost comparison of different biometric domains.

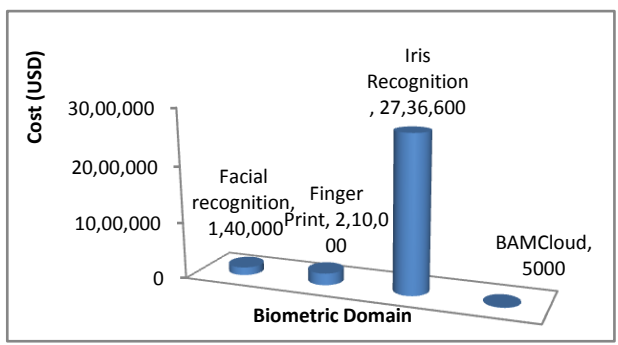

Fig. 9: Cost comparison of different biometric domains

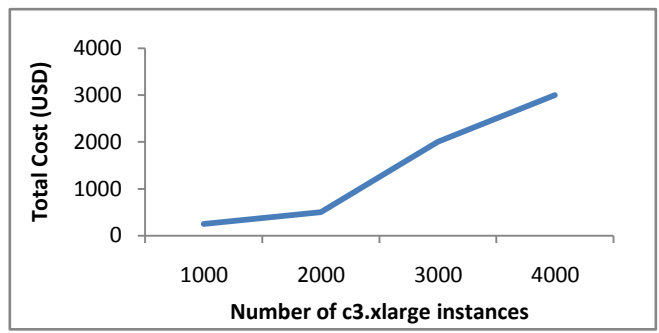

Fig. 10: Number of VMs versus total cost (in USD) on Amazon EC2

TABLE VII
BIOMETRIC DOMAINS AND THEIR COSTS

| Biometric Domain | Cost (USD) | References | Technology/ Algorithm Used |
|---|---|---|---|
| Facial recognition | 140,000 | [42] | Facial recognition Biometric hardware |
| Finger Print | 210,000 | [43] | Finger print scanner |
| Iris Recognition | 273,6600 | [44],[45] | IrisGuard |
| Signature Recognition (BAMCloud) | 5000 | – | Cloud, Hadoop, Neural Network |

Here, proposed method has been compared with three already existing biometric methods used in literature i.e. facial recognition, finger print and iris. The results show that the cost incurred for implementing the current system is lower than the methods used in literature. Thus, BAMCloud outperforms the ones existing in literature.

Table VII, furthermore shows the comparative analysis of the cost incurred in the running of these biometric systems. It is analyzed that, the implementation and running cost of iris recognition technology is the highest amongst the others, this can be attributed to the costly hardware required to capture the data for iris recognition system, whereas facial and fingerprint recognition technologies require lesser cost but the hardware cost is still implicit in these systems. Even though facial and fingerprint recognition features are now available in few high end mobile phones on the contrary signature biometric is available on all the smartphones and achieves an accuracy comparable to the traditional ones [46]. Thus, from table VII and Fig. 9, it can be concluded that BAMCloud requires least cost due to no explicit hardware requirements and high availability on all the phones.

Fig. 10 shows how the cost of the system varies with increase in demand for processing and thus increasing the number of VM instances. It should be noted that c3.xlarge on Amazon EC2 was used in the analysis and the amounts charged by Amazon EC2 for running this type of instance is 0.21 USD. From this it can be inferred that the cost for implementing the system increases linearly as the processing requirements increase and hence the number of instances required increases. Therefore, this system is viable in small, medium and large scale deployment. Since the cost increases linearly, the implementation of the biometric system can be scaled up as and when required without any exponential rise in cost as prevalent in the existing systems.

Thus, from the cost and benefit analysis of the proposed system it can be concluded that use of Cloud based mobile biometric is more cost efficient adhering to the efficiency of the system.

## VI. CONCLUSION AND FUTURE WORK

In order to meet up for the data storage and processing challenges imposed by the ever increasing demands of the biometric signature samples, BAMCloud a Cloud based mobile biometric authentication framework is proposed. This framework uses parallel algorithms for training and processing and thus, is able to handle storage and processing data set of any volume. The experimental results show that the proposed distributed data processing ALGODistSigPreprocess achieved a speedup of 10x over the other existing approaches and the training algorithm (AlgoDistTrainSample) achieved a speedup of 7x. Thus, BAMCloud gained an average speedup of 8.5x over the existing systems. The results have clearly shown that improved performance of biometric signature authentication system can be achieved using this approach. Moreover, the use of Cloud technologies offers a scalable and cost effective solution.

An analysis of the different features was also performed in order to find out the impact of different features on the system both on the distributed and sequential version of the proposed training algorithms as summarized in Table VI. The experimental results demonstrate that acceleration in X and Y direction is an important feature and has a significant impact on the system. With reported accuracy (EER 0.24) and the achieved speedup (8.5x), the system outperformed the existing systems listed in table VII. Furthermore, BAMCloud can be successfully deployed in a banking scenario where thousands of its world-wide customers are given flexibility to use their mobile device for automated authentication during internet banking using their mobile.

The proposed framework is foolproof for fraud detection also as the training data chosen for the proposed system is having sufficient number of skilled forgery examples. These frauds can be of kinds like fraud detection during enrolment phase and real time detection of identity frauds. So, in this way an attempt has been made to provide security solutions for mobile banking customers.

ACKNOWLEDGMENT

This is to acknowledge that Kashish Ara Shakil is the corresponding author of this paper. It is further acknowledged that Farhana Javed Zareen and Kashish Ara Shakil share equal contributions for the work carried out in this article.

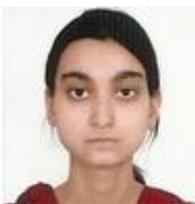
Farhana Javed Zareen, Ph.D. Scholar, Jamia Millia Islamia, Central University, New Delhi, India. She is working in the field of biometric authentication from past 3 years. She has received her bachelor's degree (2010) and master's degree (2012) in computer science from Calcutta University. Her research interests include pattern recognition, biometric authentication, image processing and machine learning.

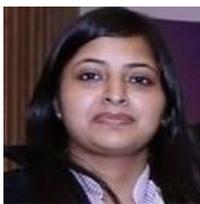
Kashish Ara Shakil has received her Bachelor's degree in Computer Science from Delhi University (2008) and has an MCA degree (2011) as well. She is currently pursuing her doctoral studies in Computer Science from Jamia Millia Islamia. Her area of interest includes database management using Cloud computing, distributed and service computing.

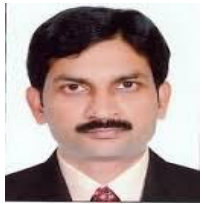
Mansaf Alam received his doctoral degree in computer Science from Jamia Millia Islamia, New Delhi in the year 2009. He is currently working as an Assistant Professor at the Department of Computer Science, Jamia Millia Islamia. He is also the Editor-in-Chief, Journal of Applied Information Science. His areas of research include Cloud database management system (CDBMS), Genetic Programming, Image Processing, Information Retrieval and Data Mining.

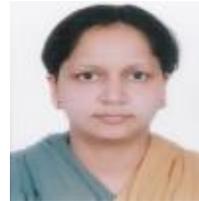
Suraiya Jabin is an Assistant professor in the Department of Computer Science, Jamia Millia Islamia, New Delhi, India. She received her Ph. D degree in 2009 from Department of Computer Science, Hamdard University India. Her research interests include Artificial Intelligence, Pattern Recognition, Soft Computing, and Biometrics.

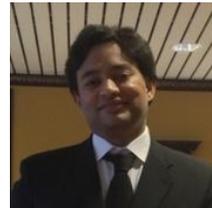
Shabih Shakeel is a postdoctoral researcher at Institute of Biotechnology and Department of Biological Sciences University of Helsinki. He received his doctoral degree (2014) from University of Helsinki, Finland. His Specialties include Transmission electron microscopy, electron cryo-microscopy, image processing and siRNA based therapeutics and its delivery system.